\documentclass[aps,preprint,groupedaddress]{revtex4}
\usepackage{amsmath,times}
\usepackage{graphicx,subfigure}

\begin{document}

\title{Neuronal Shot Noise and Brownian $1/f^2$ Behavior in the Local Field Potential}
\author{J.N. Milstein $^1$}
\altaffiliation{Corresponding author. E-mail: milstein@caltech.edu}
\author{F. Mormann $^{1,2}$}
\author{I. Fried $^{2,3}$}
\author{C. Koch $^1$}
\affiliation{$^1$ California Institute of Technology, Pasadena, Ca. 91125}
\affiliation{$^2$ Department of Neurosurgery, David Geffen School of Medicine and Semel Institute of Neuroscience and Human Behavior, University of California, Los Angeles, Ca. 90095 }
\affiliation{$^3$ Functional Neurosurgery Unit, Tel-Aviv Medical Center and Sackler Faculty of Medicine, Tel-Aviv University, 69978 Tel-Aviv, Israel}

\begin{abstract}  We demonstrate that human electrophysiological recordings of the local field potential (LFP) from intracranial electrodes, acquired from a variety of cerebral regions, show a ubiquitous $1/f^2$ scaling within the power spectrum.  We develop a quantitative model that treats the generation of these fields in an analogous way to that of electronic shot noise, and use this model to specifically address the cause of  this $1/f^2$ Brownian noise.  The model gives way to two analytically tractable solutions, both displaying Brownian noise:  1) uncorrelated cells that display sharp initial activity, whose extracellular fields slowly decay and 2) rapidly firing, temporally correlated cells that generate UP-DOWN states.
\end{abstract}


\maketitle

\section{Introduction}
Power laws appear in a large variety of settings throughout nature and often signify that there is a simple process at the origin of what appears to be a very complex phenomena.  Examples of the variety of settings in which power laws appear are the Gutenberg-Richter law for the size of earthquakes \cite{Gutenberg}, the allometric scaling laws that appear throughout biology \cite{Brown}, and Paretos's law of income distributions \cite{Pareto}.

Power laws have also been witnessed within the brain in electroencephalographic (EEG)  and magnetoencephalographic (MEG) recordings while studying a wide variety of brain function  \cite{EEGandMEG1}.  The signals recorded outside the skull by these techniques represent the global activity of a large amount of cortical and subcortical tissue and give rise to a range of exponents.  Much more local measurements of cerebral activity may be recorded by a single microelectrode.  While the emphasis of these measurements is usually focused on the spiking activity of single cells within the vicinity of the recording electrode, local field potentials (LFPs), which comprise the much slower, background of electrical activity, may also be extracted from the signal.

\begin{figure}[thp]    
\centering      
\subfigure{                   \includegraphics[width=.525\textwidth]{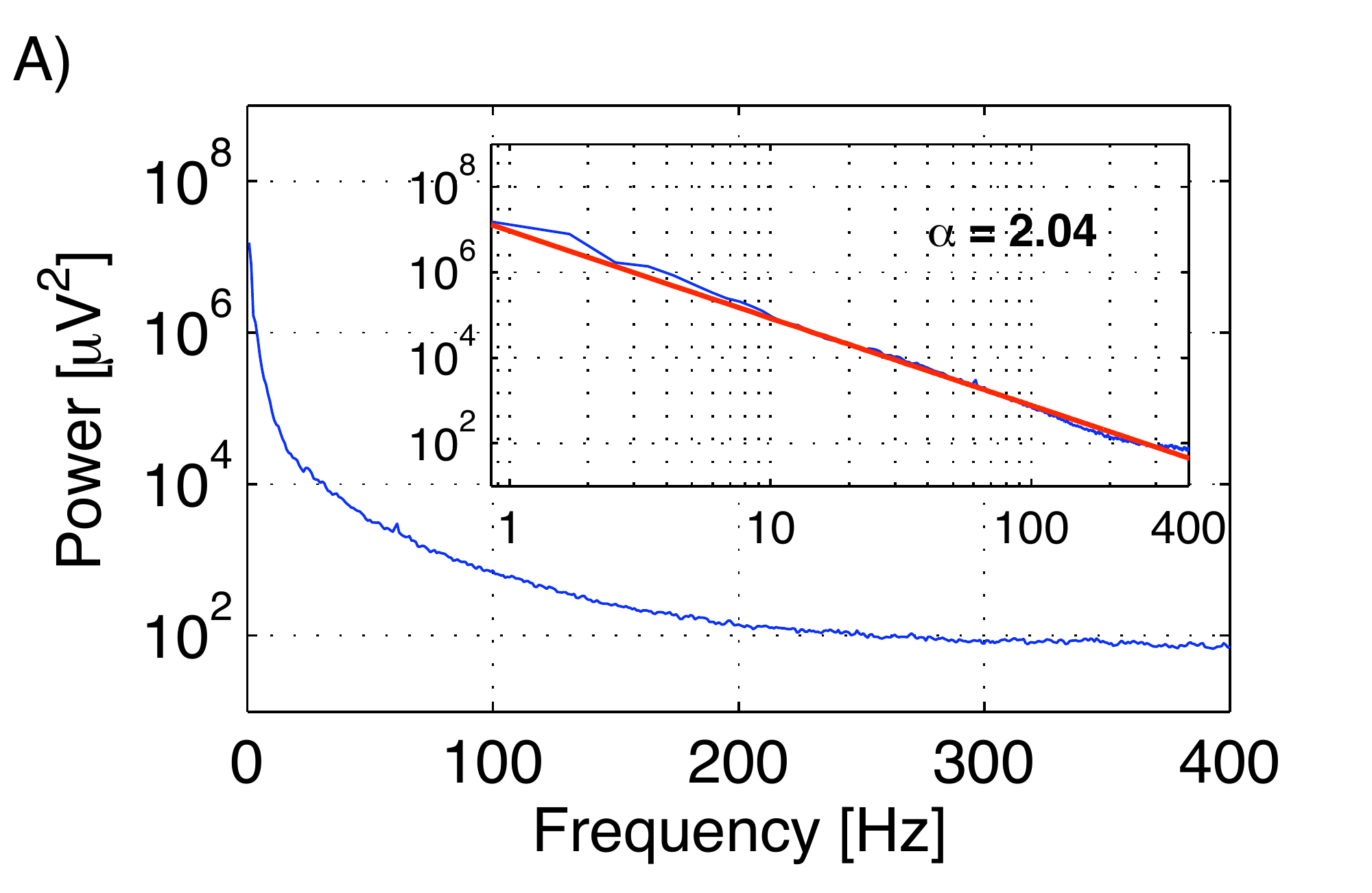}}
 \subfigure{                   \includegraphics[width=.525\textwidth]{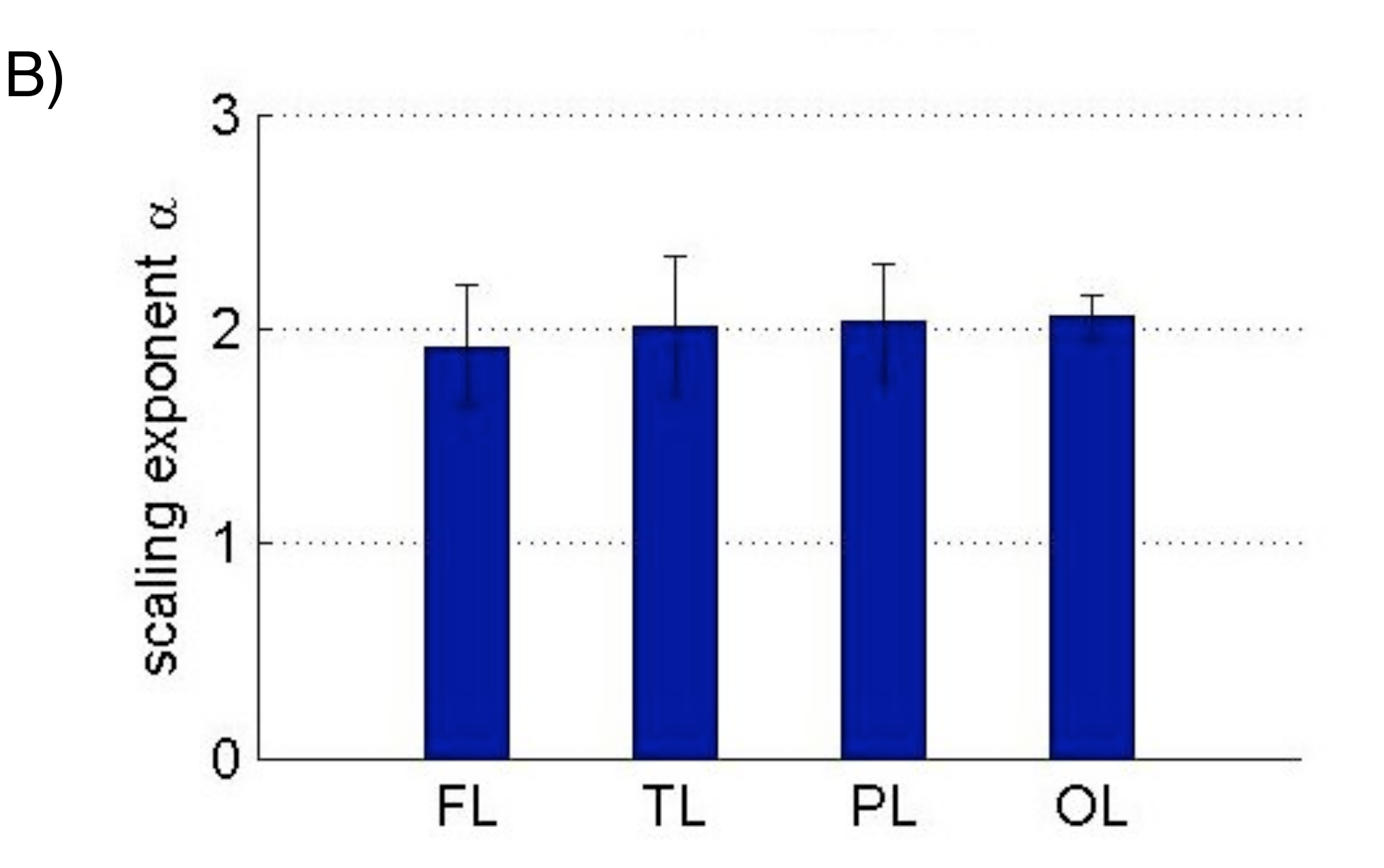}}
 \caption{Power law and scaling exponent in local field potentials recorded from the human cerebral cortex. A. Exemplary power spectrum of local field potentials recorded from a micro-wire in the temporal lobe. The scaling exponent (here $\alpha=2.04$) was determined by a linear least-square fit of the log-log power spectrum (see inset). B. Scaling exponents (mean $\pm$ stand. dev.), averaged across micro-wires for different brain regions. FL: frontal lobe; TL: temporal lobe; PL: parietal lobe; OL: occipital lobe.}
 \label{fig:humelec} 
\end{figure}

We found that electrophysiological recordings, taken from pharmacologically intractable epilepsy patients with microelectrodes implanted in a variety of cerebral areas, display a surprisingly universal $1/f^2$ power law in the frequency spectrum of LFP activity (Fig.~\ref{fig:humelec}).  A power spectrum of this sort is said to display the statistics of Brownian noise since it has the same scaling exponent as a 1D random walk.  However, it is far from clear what the underlying mechanism is that gives rise to these statistics.

To address this question, we developed a general method for modeling the LFP from what we refer to as neuronal shot noise, that allows one to study the microscopic origin (i.e., single neuron activity) of the power law dependence in the power spectrum.  We propose two quite different processes that could both give rise to the observed $1/f^2$ dependence.  The first involves the uncorrelated firing of single neurons that display very slow dendro-synaptic decay in the extracellular field which they generate.  The second possibility involves the correlated firing of a single neuron which displays either no activity (DOWN state) or very rapid spiking (UP state).  We end with a discussion of the UP-DOWN states (UDS) suggested by our model and how they compare to experimentally observed UDS within the cortex.
 \vspace{-2mm}
\section{Brownian Noise in Human Electrophysiology}
We recorded local field potentials from the cerebral cortex of 10 subjects with pharmacologically intractable epilepsy (4 males; $24-46$ years old), implanted with chronic electrodes to localize the seizure focus for possible surgical resection. Electrode locations were based exclusively on clinical criteria and were verified by MRI or by computed tomography co-registered to preoperative MRI. Each electrode probe had nine micro-wires (Platinum/Iridium, $40 {\rm \mu m}$ diameter) protruding from its tip, eight high-impedance recording channels (typically $200-400 {\rm k\Omega}$) and one low-impedance reference with stripped insulation. The differential signal from the micro-wires was amplified using a 64-channel Neuralynx system, filtered between 1 and 9000 Hz, and sampled at 28 kHz. We recorded from a total of 684 micro-wires (106 in the frontal lobe, 546 in the temporal lobe, 16 in the parietal lobe, 16 in the occipital lobe). Recordings lasted for 10 min while subjects were awake and at rest with eyes open. All studies conformed to the guidelines of the Medical Institutional Review Board at UCLA \cite{FriedNeuron}.

 For analysis, the data was down-sampled to 7 kHz using an anti-aliasing filter. The power spectral density was estimated by applying Welch's method to consecutive 5-sec segments and subsequently averaging over the entire 10 min (Fig. 1A). The scaling parameter $\alpha$ was determined as the slope of a least-square linear fit of the double-logarithmic power spectrum. To diminish the influence of amplifier roll-off, the linear fit was restricted to a frequency range of 1 to 400 Hz (Fig. 1A, inset). Figure 1B shows the scaling parameters averaged across different micro-wires for four different regions of the cerebral cortex along with their standard deviation. Note that in all four regions the scaling parameter is close to a value of $\alpha=2$, indicating a universal scaling behavior of local electrical brain activity.  To the best of our knowledge, this is the first time such a universal feature of the LFP has been reported in humans.
\vspace{-2mm}
\section{Neuronal Shot Noise}
\begin{figure}[t]
\includegraphics[width=0.8\columnwidth]{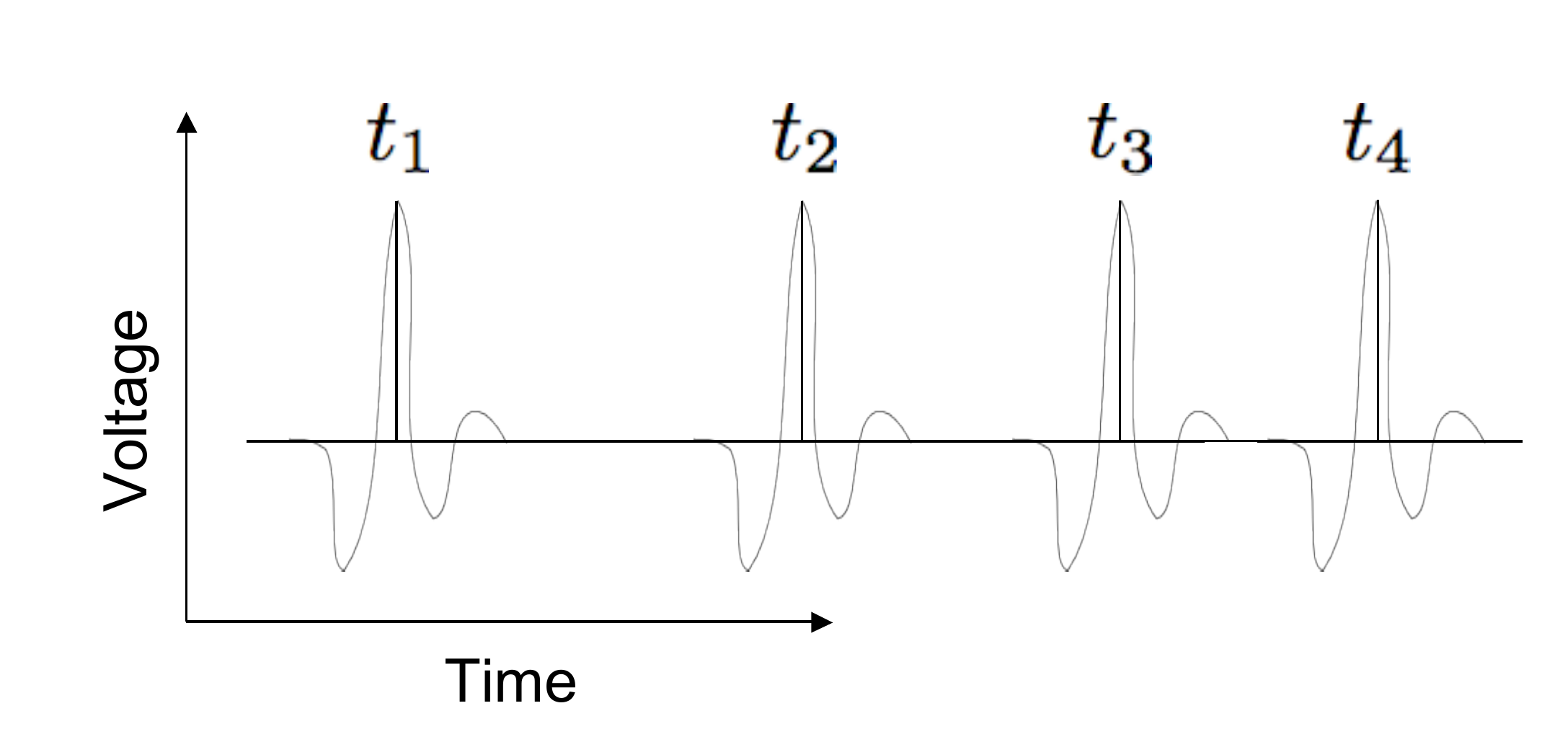}
\caption{Schematic of Eq.~\ref{shotnoise} representing neuronal shot noise.  A function $f_i(t)$, representing the extracellular field associated with the $i^{th}$ neuron, occurs at times $t_1,t_2\ldots$ governed by the statistics of $\mu_i(t)$. } 
\label{shot}
\end{figure}

The microscopic generation of the Local Field Potential (LFP) may be formulated in a similar way to that of shot noise, originally described by Schottky \cite{Schottky1} to account for the noise across an electrical resistor.  This may be seen by writing the extracellular potential $V(t)$ generated by $N$ neurons, at a given spatial location within the brain, as follows:
\begin{equation}\label{shotnoise}
V(t)=\sum_i^N\int dt' f_i(t-t')\mu_i(t'),
\end{equation}
which is exactly how one quantitatively models shot noise.  Here the function $f_i(t)$ accounts for the temporal profile of the extracellular field generated by neuron $i$ while $\mu_i(t)$ represents the activity of that neuron and may be explicitly written as $\mu_i(t)=\sum_{k_i}\delta(t-t_{k_i})$,
where $\delta(t)$ is the Dirac delta function.  From this definition, we see that the function $\mu_i(t)$ may be thought of as analogous to the spike train with firing activity occurring at times $t_{k_i}$ for neurons $i=1\ldots N$.  Note, this model does not require that the neurons generate action potentials, it only assumes a stereotyped profile $f_i(t)$ for the electrical field generated by each neuron which repeats at times $t_{k_i}$ (see Fig.~\ref{shot}).

We will assume that the relevant neural activity has reached a steady-state such that the autocorrelations \mbox{$G(\tau)=\langle V(t)V(t+\tau)\rangle$} are independent of $t$.  By the Wiener-Kinchin theorem, the autocorrelation function $G(t)$ is simply related to the power spectrum $S(\omega)$ by Fourier transformation.
From Eq.~\ref{shotnoise}, we may write the power spectrum as
\begin{equation}\label{pow}
S(\omega)=\langle|\tilde V(\omega)|^2\rangle=\sum_{i,j}\tilde f_i^*(\omega)\tilde f_j(\omega) \langle\tilde\mu_i(\omega)\tilde\mu_j(\omega)\rangle,
\end{equation}
where $\tilde V(\omega)$, $\tilde f(\omega)$, and $\tilde\mu(\omega)$ are the Fourier transforms of their respective temporal functions.
\vspace{-2mm}
\section{Biophysical Examples}
To solve for the power spectrum as written in Eq.~\ref{pow} would require us to know the location of each neuron involved in generating the LFP, the extracellular field produced by each neuron, and the decay of that field through the neuronal medium.  While we have carried out such biophysical calculations in the past for single neurons \cite{MilsteinHoltGold}, we are here only concerned  with understanding the source of the $1/f^2$ behavior of the power spectrum, not in reconstructing the LFP.  

The scale invariant nature of $S(\omega)$ greatly simplifies our problem since we need not concern ourselves with any constants that arise from the details mentioned above.  To clarify this point, let us assume that the power spectrum $S(\omega)=C_1\omega^n$, where $C_1$ is a constant.  We can solve for the coefficient of the power by plotting the log of both sides of this equation,
$\log S(\omega)=n\log\omega+\log C_1$.
The power dependence is given by the slope $n$ and is unaffected by the constant offset.  

Since the power law dependence of $S(\omega)$ is independent of overall constants, we may neglect the previously mentioned biophysical details and focus on finding solutions to Eq.~\ref{pow} that have a $1/\omega^2$ functional dependence.  In general, this problem is still quite difficult; however, there are two limits which allow a simple, analytical solution.  We will now discuss these two cases.

{\it Case I: Slow Dendro-Synaptic Decay.}
The simplest case to consider is that the spiking statistics are independent between neurons, and that the spiking of each neuron is an independent Poisson process \cite{KochBook}.  In this case 
\begin{equation}
\langle\mu_i(\omega)\mu_j(\omega)\rangle=\delta_{i,j}\bar \mu^2_i,
\end{equation}
where $\bar\mu_i$ is the average firing rate of the neuron $i$ and $\delta_{i,j}$ is the Kronecker delta function.

We can now ask when the power spectrum satisfies
\begin{equation}
S(\omega)=\sum_i \bar\mu^2_i|\tilde f_i(\omega) |^2\propto \frac{1}{\omega^2}.
\end{equation}
The solution requires $\tilde f_i(\omega)\propto 1/\omega$, whose Fourier transform is a Heaviside step function $f_i(t) \propto \theta(t)$.  This answer is a bit unrealistic since it implies that the field generated by the cell does not decay.  A more realistic solution would be to assume a form such as 
\begin{equation}
f_i(t)\propto \theta(t) e^{-\alpha t},
\end{equation}
which has Fourier transform $1/(\alpha +i\omega)$.  In the limit of slow decay, $\alpha\ll 1$, a neuron with an extracellular field of this form, firing with Poisson statistics, would give rise to Brownian noise in the LFP.

In this case, the $1/f^2$ behavior originates from the steep onset of the extracellular field.  The rise time of an action potential may occur within a fraction of a millisecond, which could account for a sharp onset, while the decay of the dendro-synaptic extracellular field may last for as long as a second \cite{Kandel}.  The functional form of the decay does not affect these results, so long as the cell takes much longer to return to baseline than it took to spike.

%
{\it Case II: UP-DOWN States.}
The second case that we consider is the limit of a sharply peaked extracellular field.  In this case, we may treat $\tilde f_i(\omega)$ as a constant $\bar f_i$, and we will assume that the activity of different neurons are either uncorrelated ($\langle\mu_i\mu_j\rangle\propto\delta_{i,j}$) or synchronous ($\langle\mu_i\mu_j\rangle\propto1$).  The spike timing of a single neuron, however, may show a temporal correlational structure.  These assumptions lead to a power spectrum
\begin{equation}
S(\omega)=\sum_{i,j}  \bar f_i \bar f_j \langle\mu_i(\omega)\mu_j(\omega)\rangle\propto \frac{1}{\omega^2}.
\end{equation}
Since all the frequency behavior is contained within the statistics of $\mu_i$, and we are assuming that all cells are active with the same statistics, we need to look for a sequence of spikes that have individual spike timing correlations of the form
\begin{equation}
\langle\mu(t+\tau)\mu(t)\rangle\propto \tau.
\end{equation}
since linear time correlations are consistent with $1/\omega^2$ frequency correlations.
This is the same linear in $\tau$ scaling as that of a 1D random walk and is at the origin of the term Brownian noise.

Since $\mu(t)$ is analogous to the spike train of each neuron, we need to formulate a binary sequence that shares the correlational structure of a random walk.  A simple way to generate a binary sequence representing white noise is to pick a random number at each timestep and then apply a threshold such that all values above the threshold are set to one, and all below to zero.  Brownian noise may be created by integrating a white noise signal.  However, it is not obvious how to apply a similar thresholding procedure to convert the resulting analog signal into a digital one \cite{foot1}.  
\begin{figure}
\includegraphics[width=0.9\columnwidth]{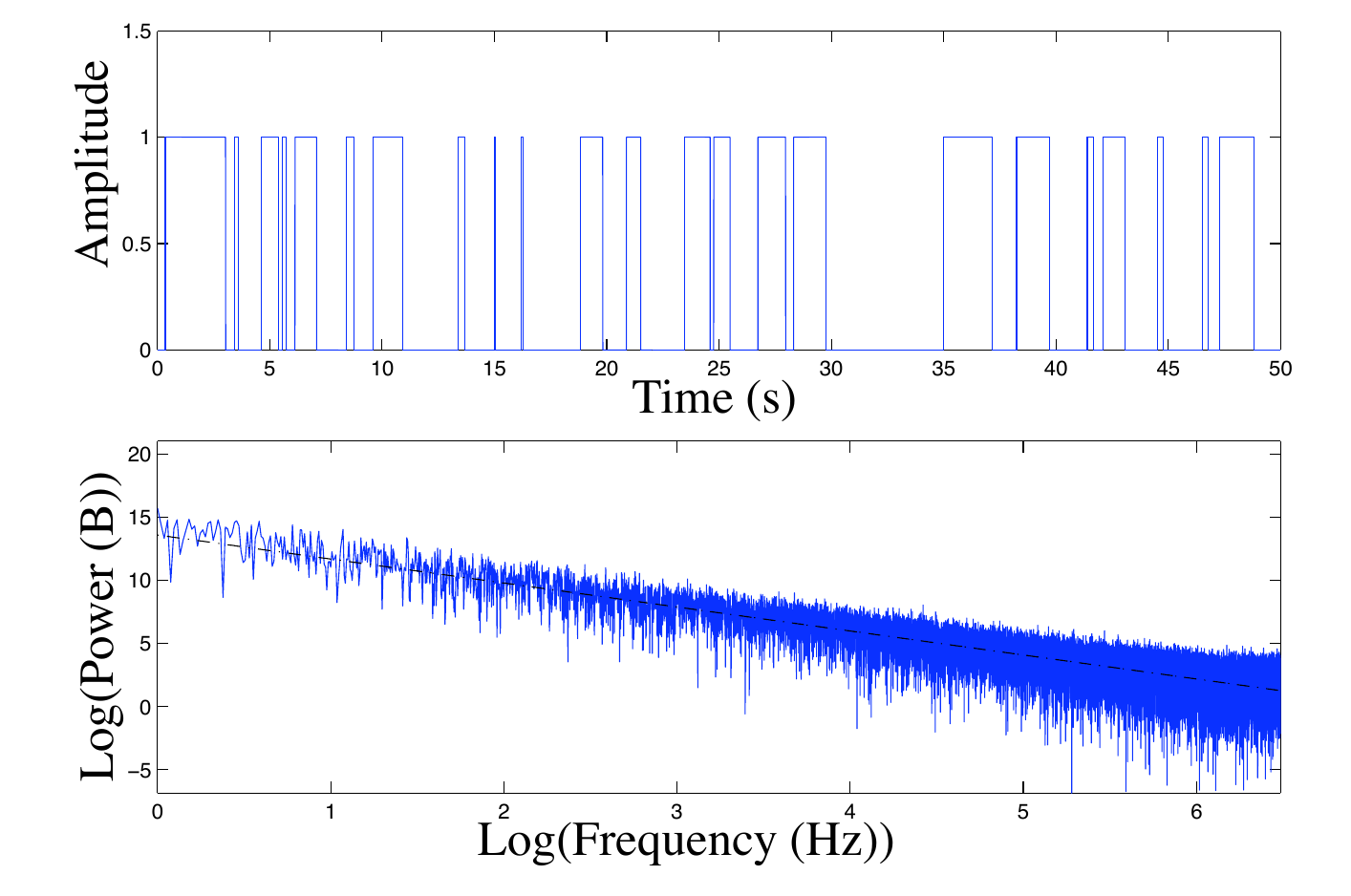}
\caption{Top: Binary sequence generated from a telegraph process.  Bottom: Power spectrum of the binary sequence confirming the $1/f^2$ behavior.  The slope of the dashed line is -2. } 
\label{brownnoise}
\end{figure}

An alternative procedure that will generate Brownian noise is given by setting up a telegraph process \cite{Gardiner}.  In this case a binary sequence is generated by constructing a two-state system,  ($0$ and $1$) where,  at each timestep, the probability of making the transition $0\rightarrow 1$ ($1\rightarrow 0$) is given by $k_+\Delta t$ ($k_-\Delta t$).  The autocorrelation function for such a process may be explicitly written as
  \begin{equation}\label{telegraph}
\langle\mu(t+\tau)\mu(t)\rangle=\left(\frac{k_+}{k_++k_-}\right)^2\left(1+\frac{k_-}{k_+}e^{-(k_++k_-)\tau}\right).
\end{equation}
In the limit of equal transition rates, $k_+=k_-$, and low probability of making a transition, $(k_++k_-)\tau\ll1$, Eq.~\ref{telegraph} reduces to
\begin{equation}
\langle\mu(t+\tau)\mu(t)\rangle\approx\frac{1}{2}(1-k\tau),
\end{equation}
which has the desired linear in $\tau$ statistics of a random walk.  

Figure \ref{brownnoise} displays a binary sequence generated by a telegraph process and the $1/f^2$ dependence of its power spectrum.  The telegraph process gives rise to periods of sustained, rapid activity followed by intervals of inactivity.  This results in collective oscillations that display a much lower frequency than the rapid firing witnessed during depolarization .  The result is a pattern of behavior reminiscent of UP-DOWN states common in cortex \cite{Steriade}.
\vspace{-2mm}
\section{Discussion}
Studies of the LFP and single neuron spiking activity, combined with current source-density analysis, suggest that LFPs are primarily the result of dendritic activity distributed over a large region of the cortex.  LFPs are therefore believed to provide a measure of the local processing and input to a given region of the brain \cite{MitzdorfLogothetis}.  

We developed a very simple model to explain our key experimental finding, a $1/f^2$ decay in the local field potential recorded from individual microelectrodes implanted into human cortex. In particular, we showed two examples of how biologically realistic, single neuron activity, parametrized by the temporal shape of the resulting extracellular fields and the statistics of cellular activity, can give rise to power law behavior within the LFP.

In Case I , we showed how a population of cells, each displaying a sharp onset of activity and a much slower decay of the extracellular field, could give rise to a Brownian power spectrum.  The time course of dendritic activity is often much longer than that of an action potential. This is in line with the above statement concerning the origin of the LFP.  However, the sharpness of the temporal onset of activity is what gives rise to the power-law behavior.  One mechanism that might account for this result would be the rapid initiation of an action potential, followed by slow dendro-synaptic decay.

Of course, this model not only assumes that the spiking statistics of each neuron is Poisson,  but that there are no correlations among  neurons.  It is not uncommon to find the firing rate of single neurons uncorrelated with the averaged behavior of the local population, however, this is not always the case \cite{Gray}.

In Case II, we explored the opposite extreme from Case I, that of rapidly firing, single neurons with linear temporal correlations.  This behavior is similar in nature to so-called UP-DOWN states seen in cortical neurons.  During periods of sleep, quiet awake behavior, or under a variety of anesthetics, spontaneous activity of neocortical neurons display slow 0.1 to 2~Hz oscillations called UP-DOWN states (UDS).  These states appear to be characteristic of slow-wave sleep \cite{IsomuraPetersenSteriade} and are thought to be involved in the consolidation of long-term memories and in learning.  The UDS of cortical pyramidal neurons are highly synchronized and may be clearly seen in LFP recordings of the cortex.  The UP states are characterized by a sustained depolarization that leads to rapid, 20-70~Hz spiking activity while the DOWN states show periods of hyper-polarized inactivity.

It should be pointed out that our recordings were performed in the awake resting state in the human cortex, whereas UDS and ultra-slow oscillations have been described only in states of low vigilance such as slow-wave sleep and anesthesia in animal studies. 
It is therefore unlikely that the power law scaling behavior observed in our recordings would be caused exclusively by the mechanisms illustrated in Case II.  Nevertheless, it is encouraging that this extreme analytical case of a $1/f^2$ power law scaling gives rise to phenomena that are actually observed in mammalian brains. 

The true origin of the $1/f^2$ behavior probably lies somewhere in-between the two limiting cases we have considered here.  Unfortunately, an analytic evaluation of Eq.~\ref{shotnoise} when there is an explicit time dependence in both the extracellular field ($f_i(t)$) and the firing statistics ($\mu_i(t)$) is, in general, difficult.  However, for a known set of $f_i(t)$ and $\mu_i(t)$, a numerical evaluation of Eq.~\ref{shotnoise} is straightforward. This formalism should, therefore, serve as a starting point in modeling power-law dependencies in the power spectrum of the LFP and in connecting this property to the underlying single neuron activity.

\acknowledgments
Funding was provided by a  fellowship from the Sloan-Swartz Foundation to J.N. Milstein, by a Marie Curie fellowship from the European Commission to F. Mormann, and by NINDS, DARPA, NSF and the Mathers Foundation. 
\vspace{-1mm}

\begin{thebibliography}{}
\vspace{-1mm}

\bibitem{Gutenberg} B. Gutenberg and C. Richter, {\it Seismicity of the Earth and Associated Phenomena}, Princeton University Press, Princeton, N.J., (1954); B. Gutenberg and C. Richter, Bull. Seismol. Soc. Am. {\bf 32}, 163 (1942).

\bibitem{Brown}  J. H. Brown and G. B. West, {\it Scaling in Biology: Santa Fe Institutes Studies in the Science of Complexity}, Oxford University Press, New York, (2000).

\bibitem{Pareto}  V. Pareto, {\it Cours d'Economie Politique}, vol. 2, Macmillan, Paris, (1897).

\bibitem{EEGandMEG1} K. Linkenkaer-Hansen, V. V. Nikouline, J. M. Palva, and R. J. Ilmoniemi, J. Neurosci. {\bf 21} 1370 (2001); E. Pareda, A. Gamundi, R. Rial, and J. Gonzalez, Neuro. Lett. {\bf 250}, 91 (1998); W. J. Freeman, M. D. Holmes, B. C. Burke, and S. Vanhatalo, Clin. Neurophys. {\bf 114}, 1053 (2003).

\bibitem{FriedNeuron} I. Fried, K. A. MacDonald, and C. L. Wilson, Neuron {\bf 18}, 753 (1997).

\bibitem{Schottky1} W. Schottky, Ann. Phys. {\bf 57}, 541 (1918).

\bibitem{MilsteinHoltGold} J. N. Milstein and C. Koch, Neural Comput. {\bf 20}, 2070 (2008); G. R. Holt and C. Koch, J. Comput. Neurosci. {\bf 6}, 169 (1999); C. Gold, D. A. Henze, C. Koch, and G. Buzsaki, J. Neurophysiol. {\bf 95}, 3113 (2006).

\bibitem{KochBook} C. Koch, {\it Biophysics of Computation: Information Processing in Single Neurons}, Oxford University Press, New York, (1999).

\bibitem{Kandel} E. R. Kandel, J. H. Schwartz, and T. M. Jessell, {\it Principles of Neural Science}, McGraw-Hill, Health Professions Division, New York, (2000).

\bibitem{foot1} For instance, one may limit the random walk to positive numbers and set a threshold.  A binary sequence is generated by adding a $0$ at each timestep that the walker remains below the threshold.  If the threshold is reached, a $1$ is added to the binary sequence, the walker is reset to the origin, and the process continues.  Unfortunately, the resetting procedure clears the memory of the random walker, and we again generate a flat, white noise spectrum.

\bibitem{Gardiner} C. Gardiner, {\it Handbook of Stochastic Methods}, Springer, New York, (1990).

\bibitem{Steriade} M. Steriade, A. Nunez, and F. Amzica, J. Neurosci. {\bf 13}, 3252 (1993).

\bibitem{MitzdorfLogothetis} U. Mitzdorf, Physiol. Rev. {\bf 65}, 37 (1985); N. Logothetis, J. Neurosci. {\bf 23}, 3963 (2003).

\bibitem{Gray} C. M. Gray and W. Singer, Proc. Natl. Acad. Sci. U S A. {\bf 86}, 1698 (1989).

\bibitem{IsomuraPetersenSteriade} Y. Isomura, A. Sirota, S. Ozen, S. Montgomery, K. Mizuseki, D. A Henze, and G. Buzsaki, Neuron {\bf 52}, 871 (2006); C. C. H. Peterson, T. T. G. Hahn, M. Mehta, A. Grinvald, and B. Sakmann, Proc. Natl. Acad. Sci. U S A. {\bf 100}, 13638 (2003); M. Steriade, I. Timofeev, and F. Grenier, J. Neurophysiol. {\bf 85}, 1969 (2001).

\end{thebibliography}
\bibliographystyle{apsrev}

\end{document}